\documentclass[11pt,a4paper]{article}
\usepackage[hyperref]{emnlp2020}
\usepackage{times}
\usepackage{latexsym}
\usepackage{multirow}
\usepackage{makecell}
\usepackage{multicol}
\usepackage{booktabs}
\usepackage{scrextend}

\usepackage{graphicx}
\usepackage{subcaption}

\usepackage[export]{adjustbox}
\usepackage{microtype}

\aclfinalcopy 

\title{Monitoring Depression Trend on Twitter\\during the COVID-19 Pandemic}

\author{Yipeng Zhang\textsuperscript{1}, Hanjia Lyu\textsuperscript{1}\thanks{\hspace{0.1cm} Equal contribution.}, Yubao Liu\textsuperscript{1}\footnotemark[1], Xiyang Zhang\textsuperscript{2}, Yu Wang\textsuperscript{1}, Jiebo Luo\textsuperscript{1}\\
  \textsuperscript{1}University of Rochester, \textsuperscript{2}University of Akron\\
  \texttt{\{yzh232,hlyu5\}@ur.rochester.edu}}

\date{}

\begin{document}
\maketitle
\begin{abstract}
The COVID-19 pandemic has severely affected people's daily lives and caused tremendous economic loss worldwide. However, its influence on people's mental health conditions has not received as much attention. To study this subject, we choose social media as our main data resource and create by far the largest English Twitter depression dataset containing 2,575 distinct identified depression users with their past tweets. To examine the effect of depression on people's Twitter language, we train three transformer-based depression classification models on the dataset, evaluate their performance with progressively increased training sizes, and compare the model's ``tweet chunk"-level and user-level performances. Furthermore, inspired by psychological studies, we create a fusion classifier that combines deep learning model scores with psychological text features and users' demographic information and investigate these features' relations to depression signals. Finally, we demonstrate our model's capability of monitoring both group-level and population-level depression trends by presenting two of its applications during the COVID-19 pandemic. We hope this study can raise awareness among researchers and the general public of COVID-19's impact on people's mental health.
\end{abstract}

\section{Introduction}
COVID-19, or the coronavirus disease of 2019, is an infectious disease that reportedly originated in China and has been spreading rapidly across the globe in 2020. It was first identified on December 31st, 2019, and was officially declared as a pandemic by the World Health Organization (WHO) on March 11th, 2020.\footnote{\label{who_covid}https://www.who.int/emergencies/diseases/novel-coronavirus-2019} As of May 29th, 2020, COVID-19 has infected 216 countries, areas or territories with over 5.7 million confirmed cases and 3.5 hundred thousand confirmed deaths\footref{who_covid}. In response to the pandemic, over 170 countries have issued nationwide closure of educational facilities\footnote{https://en.unesco.org/covid19/educationresponse}, and many governments have issued flight restrictions and stay-at-home-orders, affecting everyday lives of people around the globe.

Multiple studies have investigated the economic and social impacts of COVID-19 \citep{fernandes2020economic, baker2020covid, nicola2020socio}, but what mental impact the drastic life changes bring to people and how to quantify it at the population level are yet to be studied. Mental disorders are disturbing approximately 380 million people of all ages worldwide.\footnote{https://www.who.int/en/news-room/fact-sheets/detail/mental-disorders} Previous studies have shown that mental disorders lead to many negative outcomes including suicide \citep{inskip1998lifetime, san2019association}, but individuals who suffer from mental disorders are sometimes unwilling or ashamed to seek help \citep{yoshikawa2017factors}. Moreover, it is infeasible for psychological studies to obtain and track a large sample of diagnosed individuals and perform statistically significant numerical analysis. 

\begin{figure}[ht]
    \centering
    \includegraphics[scale= 0.145]{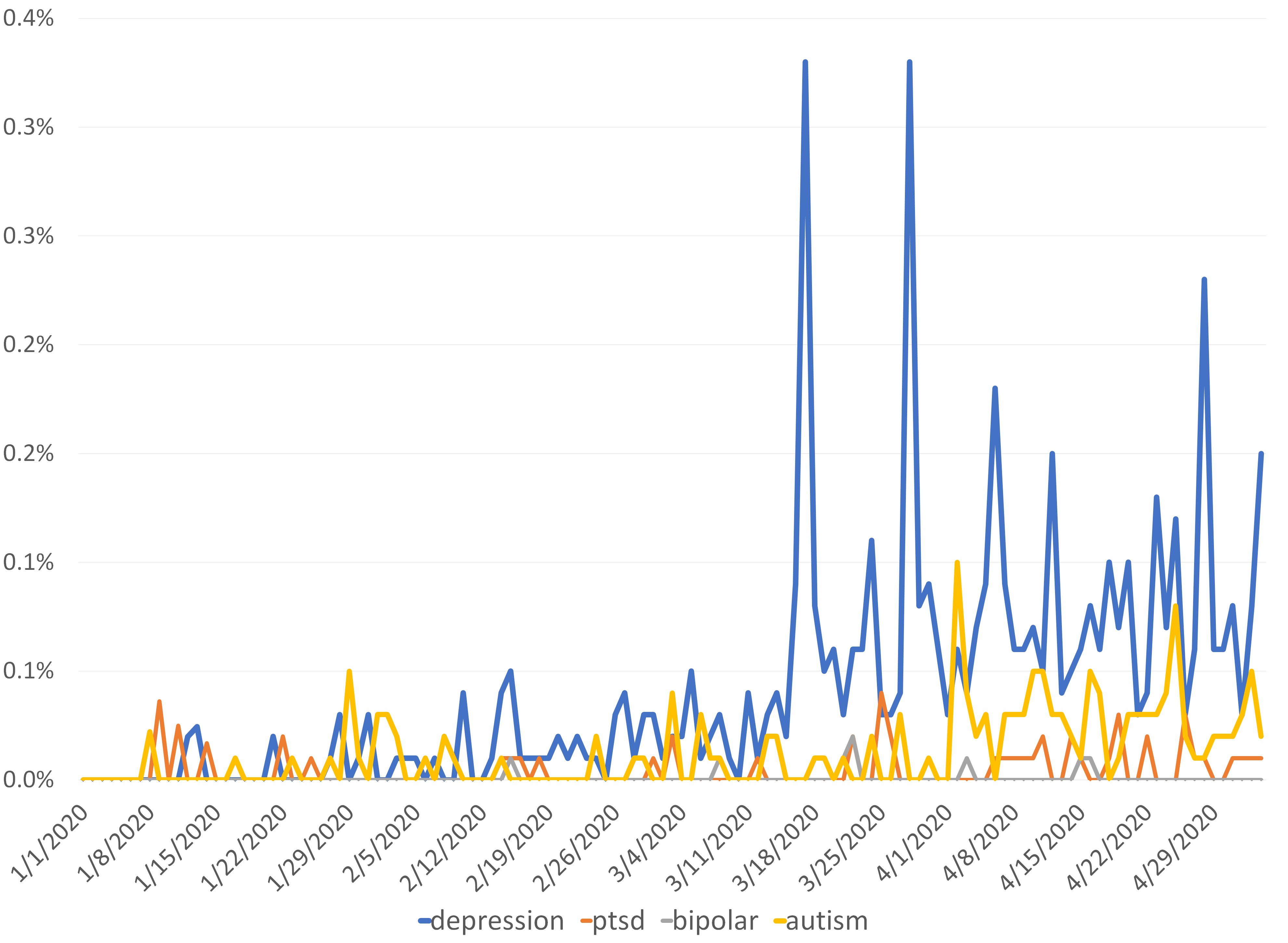}
    \vspace{-0.1cm}
    \caption{Density of Twitter coverage regarding ``depression", ``ptsd", ``bipolar disorder", and ``autism".}
    \label{fig:timeline}
\end{figure}

In the past decade, people have been increasingly relying on social media platforms such as Facebook, Twitter, and Instagram to express their feelings. We believe that social media is a resourceful medium to mine information about general public's mental health conditions. As shown in Figure~\ref{fig:timeline}, we use data from the ForSight by Crimson Hexagonplot\footnote{https://www.brandwatch.com/} to plot the word frequencies of several mental disorders on Twitter, including ``depression", ``PTSD", ``bipolar disorder", ``autism", from January 1st, 2020 to May 4th, 2020. Note that we exclude false positive tweets that contain misleading phrases such as ``economic depression" or ``great depression". We notice a rapid growth of the word frequencies of autism and depression starting from March 17th, when the pandemic spread across most of the globe\footref{who_covid}. Since depression occurs substantially more frequently compared to the other three mental disorders, we focus on understanding COVID-19's impact on depression in this study. 

Previous studies have used n-gram language models \citep{coppersmith2014quantifying}, topic models \citep{armstrong2018using}, and deep learning models such as 1-dimensional convolutional neural networks (CNN) and bidirectional long short-term memory (BiLSTM) \citep{orabi2018deep} to classify depression at the user level using Twitter data. All these works use small samples of fewer than 500 users. \citet{shen2017depression} extend previous studies by expanding the dataset to contain 1,402 depression users and using a multimodal dictionary learning approach to learn the latent features of the data. 

In this study, we create a dataset of 5,150 Twitter users, including half identified depression users and half control users, along with their tweets within past three months and their Twitter activity data. We develop a chunking and regrouping method to construct 32,420 tweet chunks, 250 words each in the dataset. Recently, transformer-based deep learning language models have achieved state-of-the-art performance in multiple language modeling tasks. We investigate the performance of some of these models, including BERT \citep{devlin2018bert}, RoBERTa \citep{liu2019roberta}, and XLNet \citep{yang2019xlnet} on our dataset. We progressively add data to our training set and notice a clear performance growth on all models, which validates the importance of our dataset. We compare the models' performance at the chunk level as well as the user level, and observe further performance gain which adds credibility to our chunking method. Furthermore, we build a more accurate classification model upon the deep learning models along with linguistic analysis of dimensions including personality, Linguistic Inquiry and Word Count (LIWC) - a well-validated psycholinguistic dictionary \citep{tausczik2010psychological}, sentiment features and demographic information. 

\citet{de2013social} demonstrate that depression prediction models can potentially be used on the population level. However, to the best of our knowledge, all Twitter user depression identification studies reviewed above focus on either tweet-level or user-level classification rather than applying the model to analyzing the mental health trends of a large population. The word frequency trend shown in Figure~\ref{fig:timeline} is apparently filled with noise and lacks plausible explanation, making it unable to reflect the actual mental health trends of the population. In this study, we present two applications to monitor the change of depression level in different groups as the disease propagates. We use latent Dirichlet allocation (LDA) and psychological theories to analyze the trend we discover. To the best of our knowledge, we are the first to apply these transformer-based models to identifying depression users on Twitter using a large-scale dataset and to monitoring the public depression trend. 

In summary, our main contributions are:
\vspace{-0.2cm}
\begin{itemize}
\item We develop an effective search method to identify depression users on Twitter. Using this method, we are able to create by far the largest dataset with 2,575 depression users along with their past tweets within three months. 
Our experiment shows that the performance of deep learning models increase as the size of the training set grows.
\vspace{-0.2cm}
\item To the best of our knowledge, we are the first to investigate the potential of transformer-based deep learning models on Twitter user depression classification.
\vspace{-0.2cm}
\item We build a tool marrying the deep learning models and psychological text analysis to further improve the classification performance. 
\vspace{-0.2cm}
\item We build a pipeline to monitor public depression trend by aggregating individuals' past tweets within a time frame using our model, and analyze the depression level trends during COVID-19, thus shedding light on the psychological impacts of the pandemic.
\end{itemize}

\section{Related Work}
The potential of machine learning models for identifying Twitter users who have been diagnosed with depression was pioneered by \citet{de2013predicting}, who analyzed how features obtained by LIWC are related to depression signals on social media and how that can be used for user-level classification on a dataset containing 171 depression users. The data was collected by designing surveys to volunteers through crowdsourcing. Following the work, \citet{coppersmith2014quantifying} used LIWC, 1-gram language model, character 5-gram model, and user's engagement on social media (user mention rate, tweet frequency, etc.) to perform  tweet-level classification on a dataset containing 441 depression users. 

The CLPsych 2015 Shared Task dataset containing 447 diagnosed depression users \citep{coppersmith2015clpsych} was published in 2015 and was favored by a wide range of studies \citep{armstrong2018using, nadeem2016identifying, jamil2017monitoring, orabi2018deep}. The data was gathered by regular expression search in tweets in combination with manual annotation. Among these studies, topic modeling was shown effective by \citet{armstrong2018using} using LDA; performance of traditional machine learning classification algorithms (decision trees, SVM, naive Bayes, logistic regression) on 1-grams and 2-grams was investigated by \citet{nadeem2016identifying}; \citet{jamil2017monitoring} used SVM on bag of words (BOW) and depression word count, along with LIWC features and NRC sentiment features; \citet{orabi2018deep} explored the performance of small deep neural architectures - 1-dimensional CNN and BiLSTM with context-aware attention on the task.

\citet{tsugawa2015recognizing} performed analysis of models using BOW, LDA, and social media engagement features on a dataset containing 81 Japanese-speaking depression Twitter users collected by crowdsourcing. One recent work \citep{shen2017depression} proposed a multimodal dictionary learning method that utilizes topic, social media engagement, profile image, emotional features to learn a latent feature dictionary that performed well on a dataset of 1402 depression users, the largest dataset used to the best of our knowledge.

\section{Data Collection and Analysis}
We use the Tweepy API to obtain 41.3 million tweets posted from March 23rd to April 18th and the information of their authors. We look for signals that can tell whether the author suffers from depression from both the text and the user profile description. Normally, Twitter users suffering from depression describe themselves as depression fighters in their descriptions. Some of them may also post relevant tweets to declare that they have been diagnosed with depression. Inspired by \citet{coppersmith2014quantifying}, we use regular expression to find these authors by examining their tweets and descriptions. Building upon their method, we further extend our regular expression search based on some patterns we notice on manually identified depression users, in pursuit of efficacy. In tweets, we search for phrases such as ``I have/developed/got/suffer(ed) from X depression'', ``my X depression'', ``I'm healing from X depression'', ``I'm diagnosed with X depression'', where X's are descriptive words such as ``severe'' and ``major''. In descriptions, we further add phrases such as ``depression fighter/sufferer/survivor'' to the regular expression list; we remove users that have ``practitioner'' and ``counselor'' in their descriptions to exclude mental health practitioners.

Once we have found the targeted Twitter users, we use the Tweepy API to retrieve the public tweets posted by these users within last 3 months by the time of posting the depression-related tweet with a max cap of 200 tweets per user. If the user is identified from the description, we limit the time scope to 3 months by April 18th. In the end, 2,575 distinct Twitter users are considered suffering from depression (DP, Depression). We randomly select another 2,575 distinct users such that depression-related terms do not appear in their past 200 tweets or descriptions as our control group. Users in this group are not considered to be suffering from depression (ND, Non-Depression). 

\subsection{Personality}

Previous psychological research shows that the Big-Five personality traits (openness, conscientiousness, extraversion, agreeableness and neuroticism) are related to depression \citep{bagby1996seasonal}. We estimate individuals' personality scores using IBM's Personality Insights service.\footnote{https://www.ibm.com/cloud/watson-personality-insights.} For each individual, we aggregate all their tweets into a single textual input and call the Personality Insights API to obtain the scores. The minimum number of words for using the API is 100 and we are able to retrieve 4,697 (91.2\%) of the 5,150 users' scores. Summary statistics are shown in Appendix~A.

\subsection{Sentiments}
\begin{figure}[ht]
    \centering
    \includegraphics[scale= 0.42]{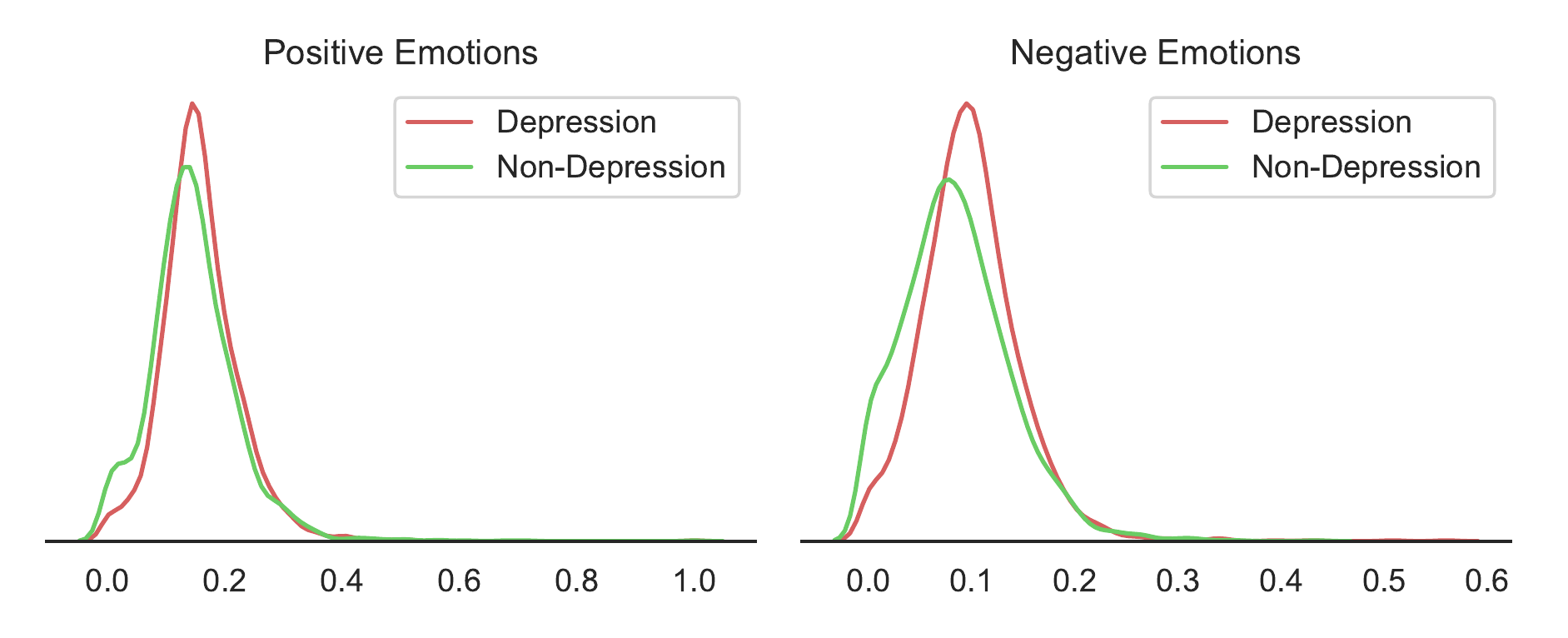}
    \caption{Compared with individuals with no depression, those with depression tend to exhibit both stronger positive and negative emotions.}
    \label{fig:sentiment}
\end{figure}

Besides personality, we hypothesize that individuals' sentiment and emotions could also reflect whether they are experiencing depression or not. We estimate individuals' sentiments using Valence Aware Dictionary and Entiment Reasoner (VADER). VADER is a lexicon and rule-based model developed by researchers from Georgia Institute of Technology \citep{hutto2014vader}. We aggregate a user’s tweets into a single chunk, apply VADER, and retrieve its scores for positive and negative emotions. In Figure \ref{fig:sentiment}, we report the distributions of positive emotions and negative emotions among DP and ND, respectively.

\subsection{Demographics}
Previous psychological studies show differences in depression rates among people of different ages and of different genders \citep{mirowsky1992age, wang2020immediate, wainwright2002childhood}. 
To estimate the age and gender of the user, we adopt the M3-inference model proposed by \citet{wang2019demographic}. The M3 model performs multimodal analysis on a user's profile image, username, and description. Following M3's structure, we label each user a binary gender label (as approximation) and a one-hot age label among four age intervals ($\leq$18, 19-29, 30-39, $\geq$40), which are then used in our fusion model. We are able to retrieve 5,059/5,150 (98.2\%) users' demographic information. 

\subsection{Linguistic Inquiry Word Count (LIWC)}
\begin{figure}[ht]
    \centering
    \includegraphics[scale = 0.5]{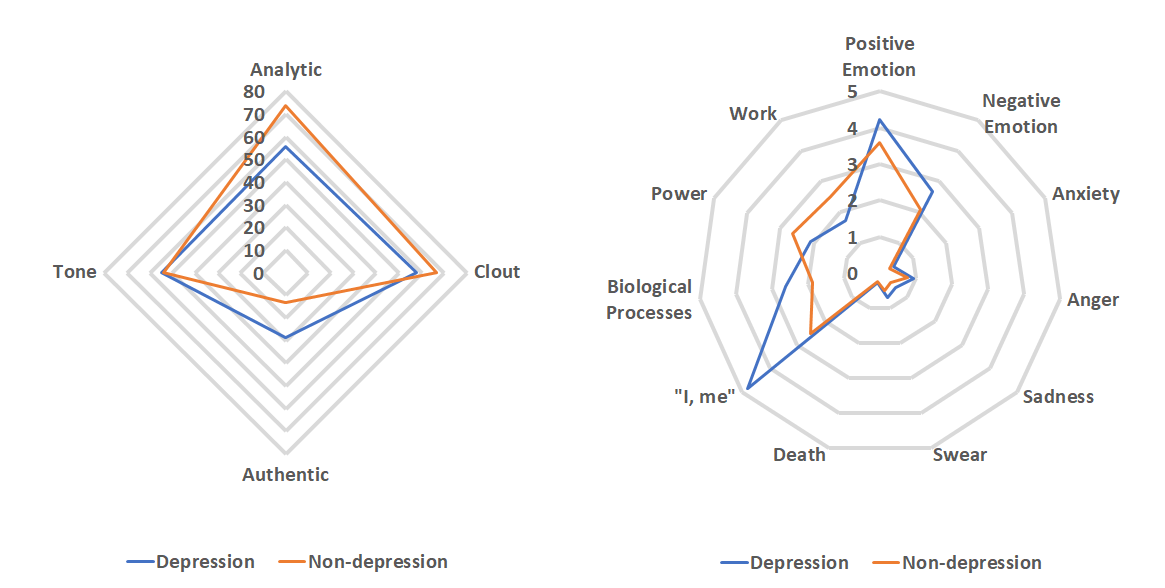}
    \caption{Linguistic profiles for the DP/ND tweets.}
    \label{fig:liwc_radar}
\end{figure}

We use LIWC to capture people's psychological states by analyzing the contents of their tweets\footnote{https://liwc.wpengine.com/how-it-works/}. We choose 8 features that were analyzed in previous works~\citep{coppersmith2014quantifying, stirman2001word, rude2004language}, as well as 7 other features that we find relevant to our study. Similar to the method conducted in ~\cite{chen2020eyes}, we apply LIWC to the concatenated tweets of individuals. Figure~\ref{fig:liwc_radar} shows the linguistic profiles for the tweets of DP and ND. Both DP and ND exhibit slightly positive tone, with negligible differences. The tweets of ND show more analytical thinking, more clout and less authentic expression than those of DP. The tweets of DP score higher in both positive and negative emotion categories than the ones of ND in which suggests a higher degree of immersion~\cite{holmes2007cognitive}. Moreover, the tweets of DP also show more anxiety, anger emotions and include more ``swear" words, consistent with the findings of~\citet{coppersmith2014quantifying}. Death-related words appear more frequently in the tweets of DP, which echoes~\citep{stirman2001word}. Similar to~\citet{coppersmith2014quantifying, stirman2001word},
we find more first-person singular in the tweets of DP. 

We also find that the tweets of DP express more sadness emotion and use words related to the biological process more frequently. While there is no clear link between biological-process-related words and depression, this finding shows that people who suffer from depression may pay more attention to their biological statuses. The ``power" score for the tweets of ND is higher which reflects a higher need for the power according to the findings of~\citet{mcclelland1979inhibited}. By comparing the ``work" scores of DP and ND, we find that the users of ND pay more attention to work-related issues as well. 

 \vspace{-0.5cm}
\subsection{Social Media Engagement}
We use the proportion of tweets with mentions, number of responses, unique users mentions, users mentions, and tweets to measure the social media engagement of each user, as did~\citet{coppersmith2014quantifying}. Our analysis details are reported in Appendix~A.

\section{Model}

\subsection{Tweet Chunking and Preprocessing}

We perform stratified random sampling on our dataset. We first sample 500 users to form our testing set. On the rest of the users, we progressively add users to the training sets and record the performance of the models trained on sets of 1,000, 2,000, and 4,650 users, respectively. All the training and testing sets have a 1-1 DP-ND ratio.

\citet{jamil2017monitoring} have shown that one single tweet does not contain enough signals to determine whether a user has depression. Thus, we concatenate consecutive tweets of the same user together to create tweet chunks of 250 words and label the chunks based on the user's label. We preprocess the text using the tweet preprocessing pipeline proposed by \citet{baziotis-etal-2017-datastories-semeval}. We adopt this method especially due to its capability of marking Twitter-specific text habits and converting them to special tokens such as ``$\langle$allcaps$\rangle$'' (capitalized words), ``$\langle$elongated$\rangle$'' (repeated letters), ``$\langle$repeated$\rangle$'' (repeated words), etc. For example, 
\begin{quote}
    YESSSSS, I love it so so much!!!
\end{quote}
after preprocessing will be in the form of
\begin{quote}
    Yes $\langle$allcaps$\rangle$ $\langle$elongated$\rangle$, I love it so $\langle$repeated$\rangle$ much! $\langle$elongated$\rangle$
\end{quote}
After chunking and preprocessing, on average, each user has 6-7 text chunks, making the actual sizes of the 4,650-user train-val set and the 500-user testing set to be 29,315 and 3,105, respectively. The preprocessed tweet chunk datasets are then passed to deep learning models for training. 

\subsection{Deep Learning Models}
We use deep learning models to perform chunk-level classification. We set up 2 baseline models, multi-channel CNN and bidirectional LSTM with context-aware attention (Attention BiLSTM), as described in \citet{orabi2018deep}. We use the pretrained  GloVe embedding (840B tokens, 300d vectors)~\citep{pennington2014glove} augmented with the special tokens added during preprocessing. The embedding weights are further trained jointly with the model. We also train three representative transformer-based sequence classification models - BERT, RoBERTa and XLNet - with their own pretrained tokenizers augmented with the special tokens for tokenization. We choose to use the base models for all of them since we find no noticeable performance gains using their larger counterparts. 

\subsection{Signal Fusion}
We run the models on all the tweet chunks of the same user and take average of the confidence scores to get the user-level confidence score. There are 4,163/4,650 (89.5\%) users remaining in the training set and 446/500 (89.2\%) in the testing set whose entire features are retrievable. We then pass different combinations of user-level scores (personality, VADER, demographics, engagement, LIWC, and average confidence) to machine learning classification algorithms including random forest, logistic regression, and support vector machines (SVM) provided by the scikit-learn library.\footnote{https://scikit-learn.org/stable/} We only use the explainable LIWC features mentioned in the data collection section for training the classifiers.

\subsection{Training Details}
During training, we randomly split the train-val set to training and validation sets with a ratio of 9:1. We record the models' performances on the validation set after each epoch and keep the model with the highest accuracy and F1 scores while training until convergence. Hyper-parameters and other experimental settings are described in Appendix~B.

\begin{table*}[t]
\centering
\small
\begin{tabular}{l|l|lllll}
\hline
\textbf{Model} & \textbf{Train-Val Set} & \textbf{Accuracy} & \textbf{F1} & \textbf{AUC} & \textbf{Precision} & \textbf{Recall} \\
\hline
Attention BiLSTM & 
\begin{tabular}{@{}c@{}}
1k users \\ 2k users \\ 4.65k users \end{tabular} &
\begin{tabular}{@{}c@{}}
70.7\ \\ 70.3 \\ 72.7 \end{tabular} &
\begin{tabular}{@{}c@{}}
69.0 \\ 68.3 \\ 71.6 \end{tabular} &
\begin{tabular}{@{}c@{}}
76.5 \\ 77.4 \\ 79.3 \end{tabular} &
\begin{tabular}{@{}c@{}}
70.9 \\ 70.7 \\ 72.1 \end{tabular} &
\begin{tabular}{@{}c@{}}
67.3 \\ 66.1 \\ 71.1 \end{tabular}
\\
\hline
CNN & 
\begin{tabular}{@{}c@{}}
1k users \\ 2k users \\ 4.65k users \end{tabular} &
\begin{tabular}{@{}c@{}}
71.8 \\ 72.8 \\ 74.0 \end{tabular} &
\begin{tabular}{@{}c@{}}
72.6 \\ 74.5 \\ 70.9 \end{tabular} &
\begin{tabular}{@{}c@{}}
77.4 \\ 80.3 \\ 81.0 \end{tabular} &
\begin{tabular}{@{}c@{}}
72.7 \\ 72.2 \\ 77.4 \end{tabular} &
\begin{tabular}{@{}c@{}}
72.6 \\ 76.9 \\ 68.9 \end{tabular}
\\
\hline
BERT & 
\begin{tabular}{@{}c@{}}
1k users \\ 2k users \\ 4.65k users \end{tabular} &
\begin{tabular}{@{}c@{}}
72.7 \\ 75.7 \\ 76.5 \end{tabular} &
\begin{tabular}{@{}c@{}}
74.4 \\ 76.3 \\ 77.5 \end{tabular} &
\begin{tabular}{@{}c@{}}
79.8 \\ 82.9 \\ 83.9 \end{tabular} &
\begin{tabular}{@{}c@{}}
72.0 \\ 76.1 \\ 76.3 \end{tabular} &
\begin{tabular}{@{}c@{}}
76.9 \\ 75.7 \\ 78.8 \end{tabular}
\\
\hline
RoBERTa & 
\begin{tabular}{@{}c@{}}
1k users \\ 2k users \\ 4.65k users \end{tabular} &
\begin{tabular}{@{}c@{}}
74.4 \\ 75.9 \\ 76.2 \end{tabular} &
\begin{tabular}{@{}c@{}}
75.7 \\ 77.9 \\ \textbf{78.0} \end{tabular} &
\begin{tabular}{@{}c@{}}
82.0 \\ 83.2 \\ 84.1 \end{tabular} &
\begin{tabular}{@{}c@{}}
74.2 \\ 73.8 \\ 74.4 \end{tabular} &
\begin{tabular}{@{}c@{}}
77.3 \\ \textbf{82.5} \\ 81.9 \end{tabular}
\\
\hline
XLNet & 
\begin{tabular}{@{}c@{}}
1k users \\ 2k users \\ 4.65k users \end{tabular} &
\begin{tabular}{@{}c@{}}
73.7 \\ 74.6 \\ \textbf{77.1} \end{tabular} &
\begin{tabular}{@{}c@{}}
75.1 \\ 76.8 \\ 77.9 \end{tabular} &
\begin{tabular}{@{}c@{}}
80.7 \\ 82.6 \\ \textbf{84.4} \end{tabular} &
\begin{tabular}{@{}c@{}}
73.2 \\ 72.6 \\ \textbf{77.5} \end{tabular} &
\begin{tabular}{@{}c@{}}
77.2 \\ 81.5 \\ 78.3 \end{tabular}
\\
\hline
\end{tabular}
\caption{\label{table:chunk-level}
Chunk-level performance (\%) of all 5 different models 
using training-validation sets of different sizes.
}
\end{table*}

\section{Experimental Results}

\subsection{Chunk-Level Classification}

In Table~\ref{table:chunk-level}, we report our classification results at the chunk level on the testing set. Our evaluation metrics include accuracy, F1 score, AUC, precision and recall.\footnote{We use 0.5 as the threshold when calculating the scores.} One immediate observation is that regardless of the model type, the classification performance improves as we increase the size of our train-val set. This shows that for building depression classification models it is imperative to have a large number of training samples. At the same time, it also confirms that the larger number of training samples in our experiments is indeed an advantage. 

Another observation is the performance gain of transformer-based models over BiLSTM and CNN models. The CNN model slightly outperforms BiLSTM, which replicates the findings of \citet{orabi2018deep}. We observe BERT, RoBERTa and XLnet invariably outperform BiLSTM and CNN regardless of the size of our training set. In particular, the XLNet model records the best AUC as well as accuracy of all models when trained with our full training set.

\subsection{User-Level Classification}

\begin{table}[ht]
\centering
\small
\begin{tabular}{l|lll}
\hline \textbf{Features} & \textbf{Accuracy} & \textbf{F1} & \textbf{AUC}\\ \hline
VADER & 54.9 & 61.7 & 54.6 \\
Demographics & 58.7 & 56.0 & 61.4 \\
Engagement & 58.7 & 62.3 & 61.7 \\
Personality & 64.8 & 67.8 & 72.4 \\
LIWC & 70.6 & 70.8 & 76.0 \\
V+D+E+P+L & 71.5 & 72.0 & 78.3 \\
XLNet & 78.1 & 77.9 & 84.9 \\
All (Rand. Forest) & 78.4 & 78.1 & 84.9 \\
All (Log. Reg.) & 78.3 & 78.5 & \textbf{86.4} \\
All (SVM) & \textbf{78.9} & \textbf{79.2} & 86.1 \\
\hline
\end{tabular}
\caption{\label{table:user-level} User-level performance (\%) using different features. We use SVM for classifying individual features.}
\end{table}

 Next, we report our experiment results at the user level. Since XLNet trained on the 4,650-user dataset outperforms other models, we take it for user-level performance comparison. Our experimental results demonstrate a huge increase on the user-level scores of XLNet shown in Table~\ref{table:user-level} compared to the chunk-level score shown in Table~\ref{table:chunk-level}. This indicates that more textual information of a user yields more reliable results on determining whether the user has depression. Building on the user-level XLNet scores, we further include VADER (V), demographics (D), engagement (E), personality (P), and LIWC (L) scores as signals. We first use all features and compare the performance of random forest, logistic regression, and SVM. We notice that SVM achieves the best scores on accuracy and F1, slightly surpassing logistic regression. Thus, we use SVM for testing the performance when using part of the features collected. 

The results are shown in Table~\ref{table:user-level}. Results have shown that using VADER, demographics, and social media engagement features alone does not help the classification by much. Classifiers using personality features and LIWC features perform relatively better. We then use all these five feature groups and obtain a better result (accuracy 71.5\%, F1 72.0\%). However, the classifier is still outperformed by XLNet, showing that the transformer-based models indeed works better on depressive Twitter text modeling compared with other approaches. We further increase the classifier's performance by using all the features, namely, VDEPL features and the averaged XLNet confidence score; the performance of the three ML algorithms does not vary much and the SVM classifier achieves the best accuracy (78.9\%) and F1 (79.2\%) scores.

\begin{figure}[ht]
    \centering
    \includegraphics[scale=0.38]{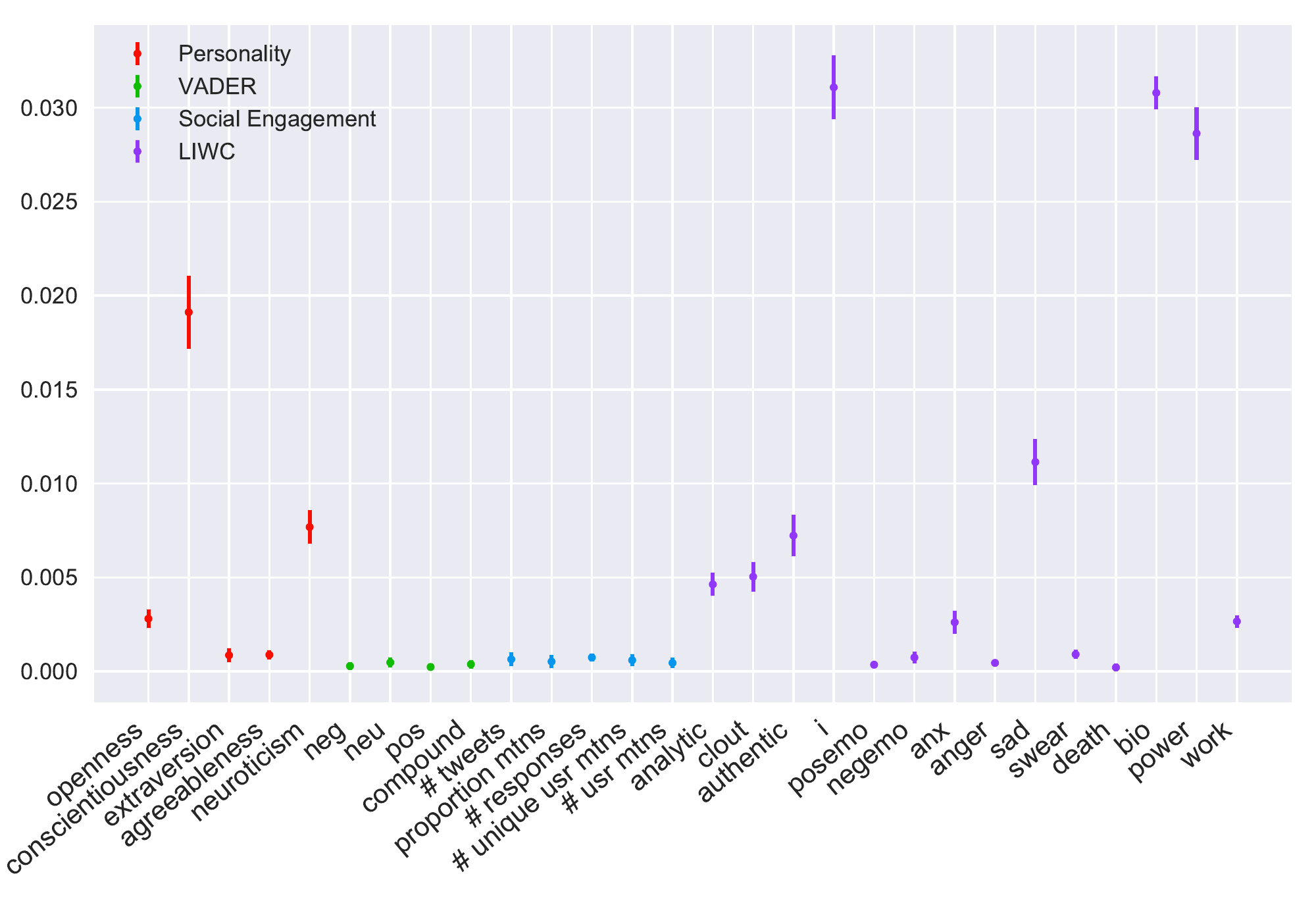}
    \caption{Permutation importance of different features.}
    \label{fig:importance}
\end{figure}

In an attempt to investigate what specific textual features besides those extracted by XLNet have the most impact on depression classification, we calculate the permutation feature importance~\citep{altmann2010permutation} on the trained random forest classifier using the VEPL features with 10 repeats. The importance scores of individual features are shown in Figure~\ref{fig:importance}. Among the LIWC features, ``i'', ``bio'', ``power'', ``sad'', ``authentic'', ``clout'' and ``analytic'' are shown to be important in classification. Among the five personality features, ``conscientiousness'' and ``neuroticism'' are shown to be closely related to depression cues. We do not observe a strong relation between VADER sentiment features or social media engagement features and the depression signals. As for the LIWC sentiment features, only ``sad'' and ``anxiety'' are shown to be relatively important; we hypothesize that the ideas a person tries to express on Twitter are more related to depression cues than his/her sentiments, which are likely to have only short-term effects.

\section{Applications}
In this section, we report two COVID-19 related applications of our XLNet based depression classifier: (a) monitoring the evolution of depression level among the depression group and the non-depression group, and (b) monitoring the depression level at the U.S. country level and state level during the pandemic.

\begin{figure*}[t]
    \centering
    \begin{subfigure}[b]{0.455\textwidth}
        \centering
        \includegraphics[width=\textwidth]{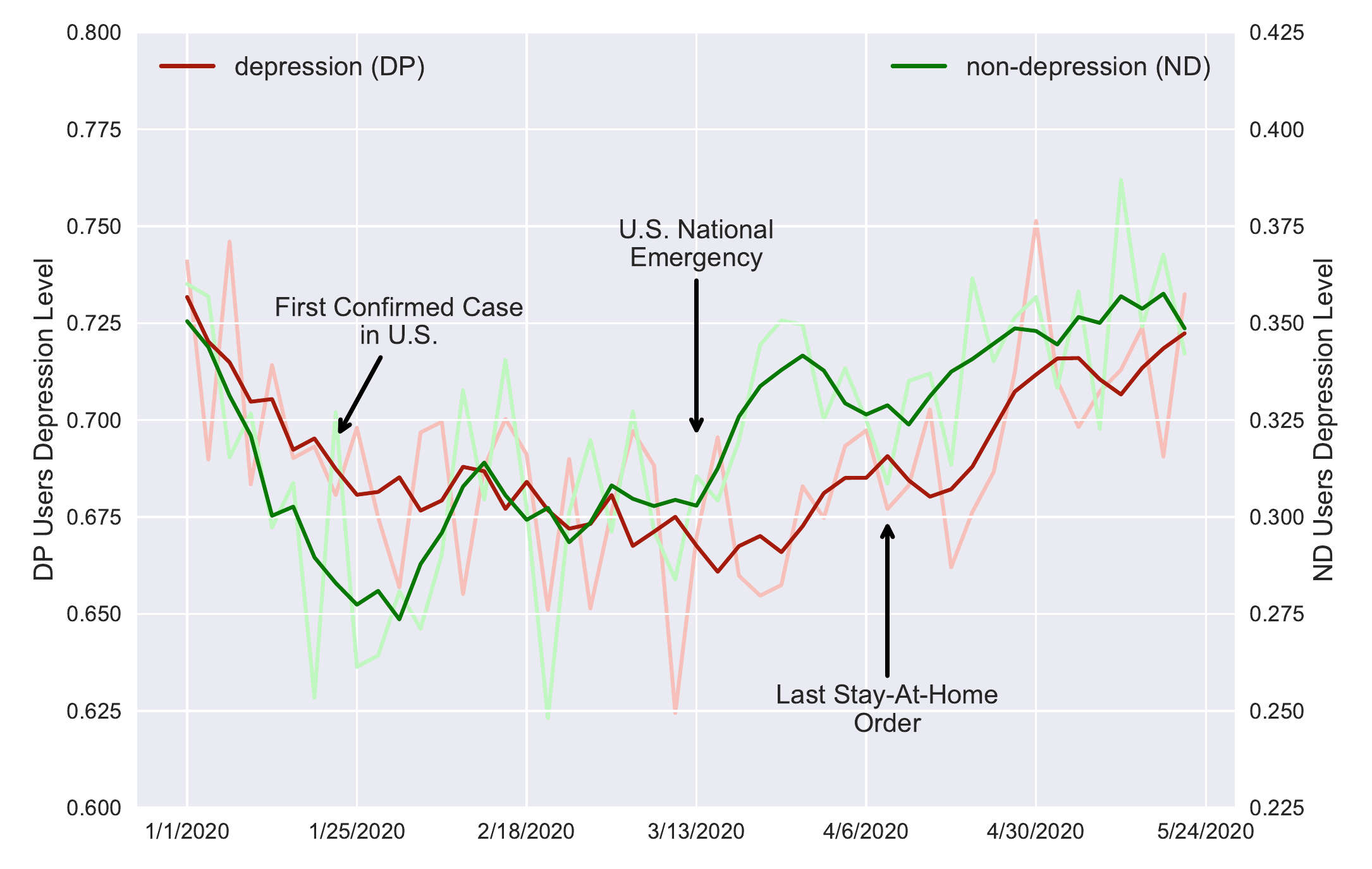}
        \caption[]%
        {{\small DP-ND trends}}    
        \label{fig:dpnd_trend}
    \end{subfigure}
    \hfill
    \begin{subfigure}[b]{0.455\textwidth}  
        \centering 
        \includegraphics[width=\textwidth]{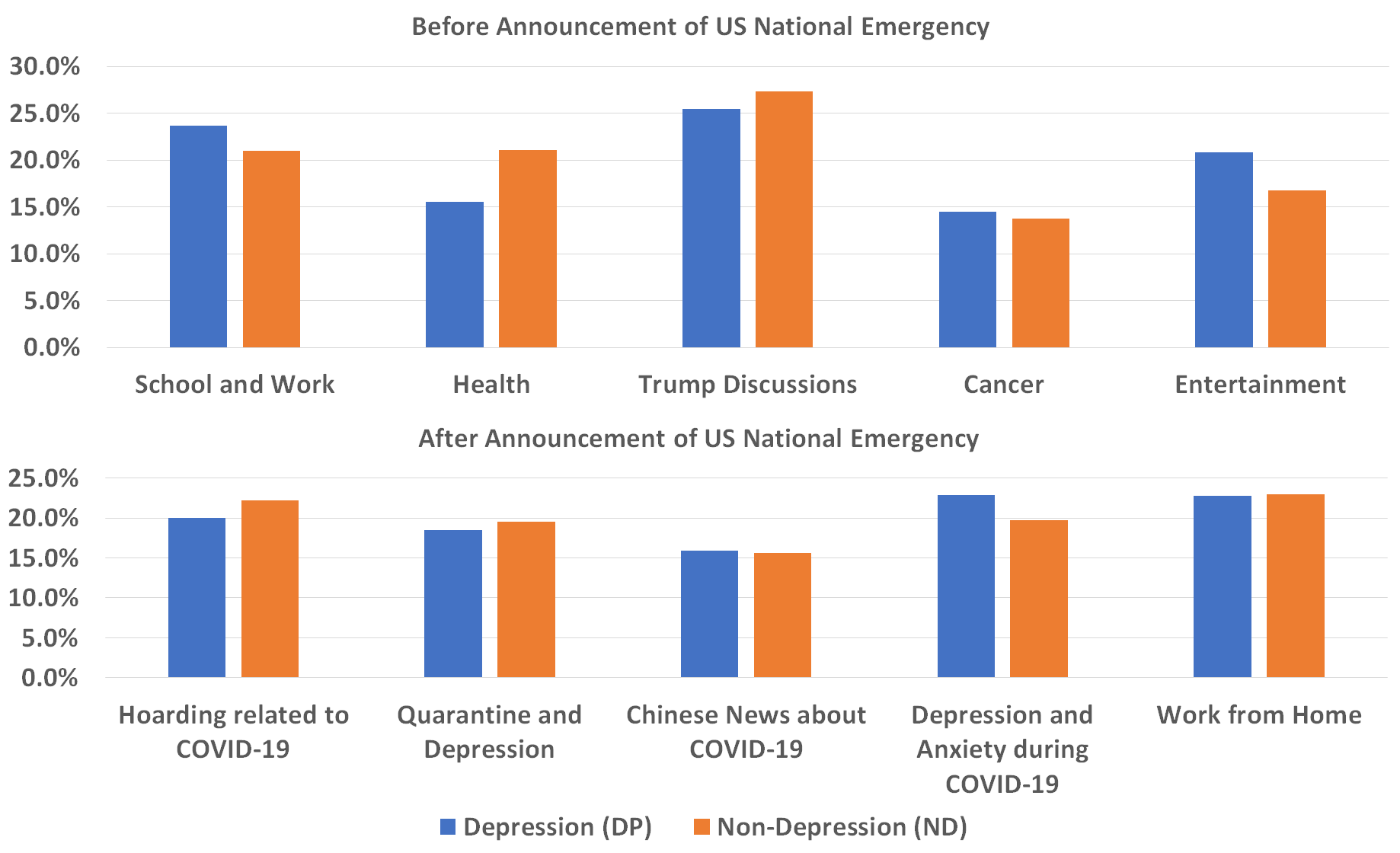}
        \caption[]%
        {{\small Percentage of DP-ND topics}}    
        \label{fig:dp_nd_lda}
    \end{subfigure}
    \vskip 0.1cm
    \begin{subfigure}[b]{0.42\textwidth}   
        \centering 
        \includegraphics[width=\textwidth]{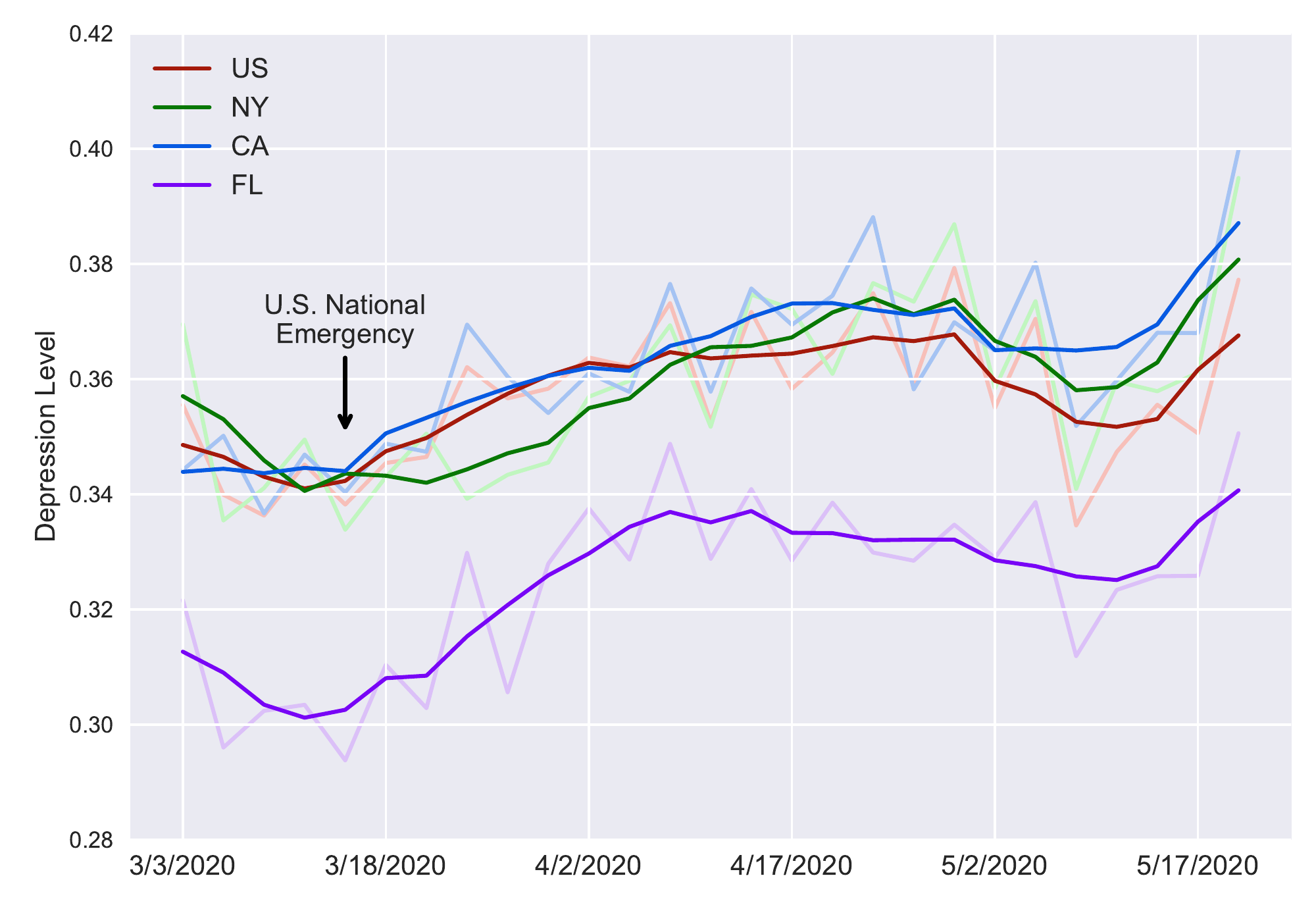}
        \caption[]%
        {{\small State-level trends}}
        \label{fig:covid_trend}
    \end{subfigure}
    \hfill
    \begin{subfigure}[b]{0.455\textwidth}   
        \centering 
        \includegraphics[width=\textwidth]{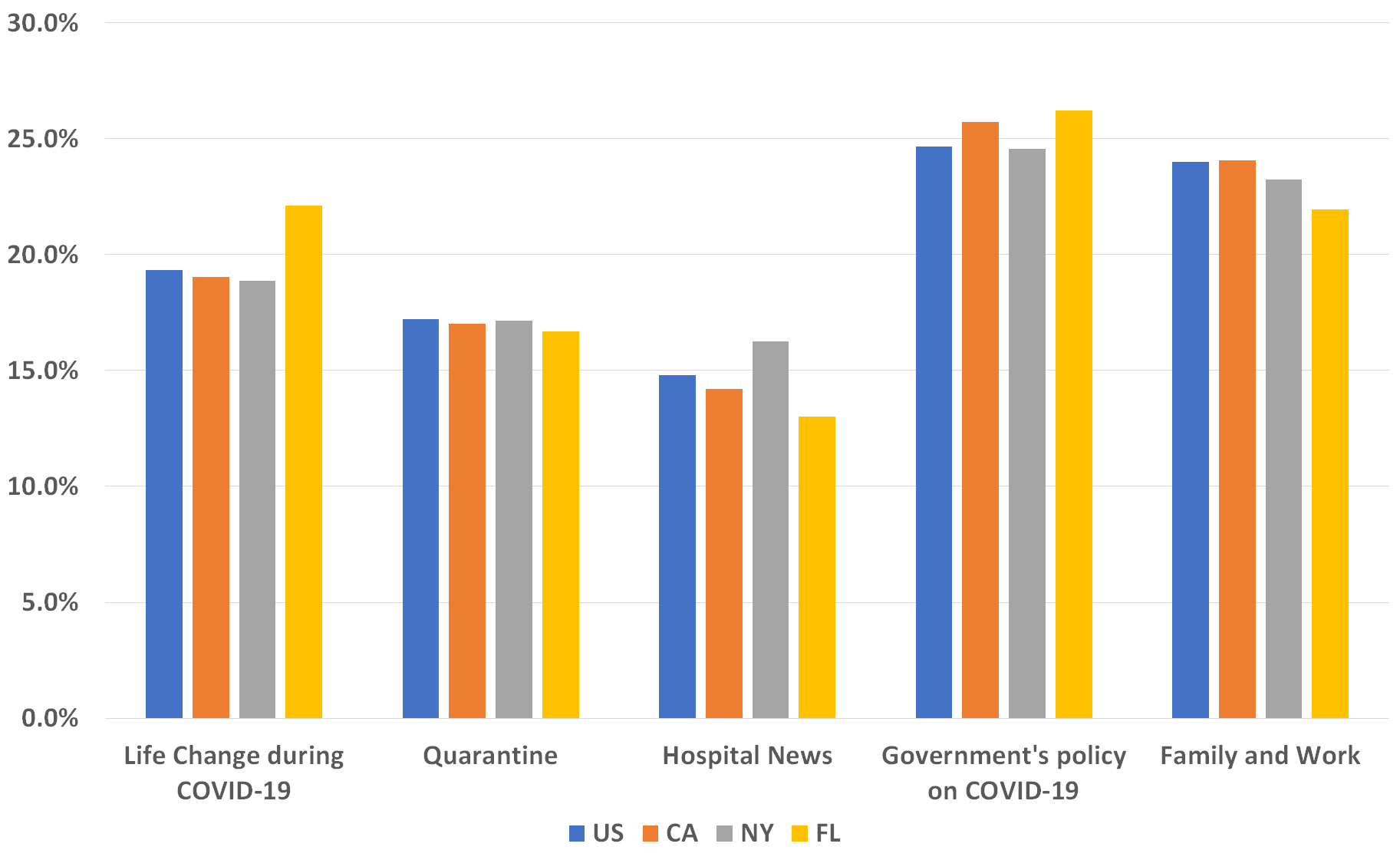}
        \caption[]%
        {{\small Percentage of State-level topics}}    
        \label{fig:state_level_lda}
    \end{subfigure}
    \vspace{-0.2cm}
    \caption[]
    {\small (a) Aggregated depression level trends of DP users and ND users from January 1st, 2020 to May 22nd, 2020. We use different y-axes for the 2 groups in order to compare them side by side. (b) Topics of DP and ND before and after the announcement of the U.S. National Emergency. (c) Aggregated depression level trends of U.S., NY, CA, and FL from Mar 3rd, 2020 to May 22nd, 2020. (d) Top 5 topics (state-level) after the announcement of the U.S. National Emergency.} 
    \label{fig:application}
\end{figure*}
\vspace{-0.2cm}

\subsection{Depression Monitoring on DP/ND Group}

We take the 500 users from the testing set along with their tweets from January 1st, 2020 to May 22nd, 2020. We concatenate a user's tweets consecutively from January 1st one by one until reaching 250 words and label this chunk's date as the date of the author posting the tweet that is in the middle of the chunk. We group three days into a bin from January 1st and assign the chunks to the bins according to the labeled date. We run the XLNet model on the preprocessed tweet chunks and record the confidence scores. We trim the upper and lower 10\% of the data to reduce the skew in the score distribution. We then take the mean of the scores for each time bin and plot the depression trend shown in Figure~\ref{fig:dpnd_trend}. We further take a moving average of 5 time bins to smooth the curves.

We mark three important time points on the plot - the first confirmed case of COVID-19 in the U.S. (January 21st), U.S. National Emergency announcement (March 13th), and the last stay-at-home order issued (South Carolina, April 7th). In January, both groups experience a drop in depression scores. This may be caused by the fact that people's mood usually hits its lowest in winter~\citep{thompson1988comparison}. From the day when there was the first confirmed case in the U.S. to the day of the announcement of U.S. National Emergency, the trends of DP and ND are different. The depression level of DP goes down slightly while the depression level of ND goes up. Aided by psychological findings, we hypothesize that depressive users are less affected by negative events happening in the outside world because they focus on their own feelings and life events, since (a) they are mostly affected by negative events that threaten them directly \citep{yue2016meta}, and (b) more interactions with the outside world give them more negative feedback \citep{winer2016reward}. Moreover, the depression levels of DP and ND both increase after the announcement of the U.S. National Emergency. 

To better understand the trend, we apply the LDA model to retrieve the topics before and after the announcement of the U.S. National Emergency. Each chunk of the tweets is assigned 5 weights for the 5 topics respectively. We label the topic of the highest weight as the dominant topic of this chunk of the tweets, and count the frequency of each topic shown in Figure~\ref{fig:dp_nd_lda}. Details about the keywords of the topics are reported in Appendix~C. Before the announcement, the 2 most frequent topics of DP and ND are the discussions about U.S. President Donald Trump and about school and work. The third most frequent topic of ND is about health, while that of DP is about entertainment. This supports the difference of the depression level trends of two groups. After the announcement of the U.S. National Emergency, the most frequent topic of DP is depression and anxiety during COVID-19, while this is the third frequent topic of ND. Further, all the 5 topics of each group are about COVID-19. This shows that when people mostly talk about COVID-19, depression signals rise for both groups.

\subsection{Aggregated Depression in COVID-19}

To investigate country- and state-level depression trend during COVID-19, we randomly sample users who have U.S. state locations stated in their profiles and crawl their tweets between March 3rd, 2020 and May 22nd, 2020, the period right before and after the U.S. announced National Emergency on March 13th. Using the same logic as in Section 6.1, we plot the change of depression scores of all geo-located 9,050 users as the country-level trend. For state-level comparison, we plot the aggregated scores of three representative states - economical center New York on the East Coast that is highly affected by the virus, tech center California on the West Coast that is also struck hard by the virus, and the less affected tourism center Florida in the southeast. Each selected state has at least 550 users in the dataset to validate our findings. Their depression levels are shown in Figure~\ref{fig:covid_trend}.

The first observation of the plot is that depression scores of all three states and the U.S. behave similarly during the pandemic; they experience a decrease right before the National Emergency, a steady increase after that, a slight decrease past April 23rd, and another sharp increase after May 10th. We also notice that the overall depression score of Florida is significantly lower than the U.S. average as well as the other two states. Since Florida has a lower score both before and after the virus breakout, we hypothesize that it has a lower depression level overall compared to the average U.S. level irrespective of the pandemic.

We calculate the topics at the state level after the announcement of the U.S. National Emergency. As shown in the Figure~\ref{fig:state_level_lda}, the most frequent topic is the government's policy on COVID-19. California and Florida are the states that pay relatively more attention to this topic compared to the U.S. average and New York. Florida also talks more about the life change during COVID-19. Another finding is that people in New York talk more about the hospital news, likely because the state contains the majority of cases in the country by May 22nd \footnote{https://www.cdc.gov/coronavirus/2019-ncov/cases-updates/cases-in-us.html}.

\section{Conclusion}
COVID-19 has infected over five million people globally, virtually brought the whole world to a halt. During this period, social media witnessed a spike in depression terms.
Against this backdrop, we have developed transformer-based models trained with by far the largest dataset on depression. We have analyzed our models' performance in comparison to existing models and verified that the large training set we compile 
is beneficial to improving the models' performance. We further show that our models can be readily applied to the monitoring of stress and depression trend of targeted groups over geographical entities such as states. We notice significant increases in depression signals as people talk more about COVID-19. We hope researchers and mental health practitioners find our models useful and this study can raise the awareness of the mental health impacts of the pandemic.

\bibliography{emnlp2020}
\bibliographystyle{acl_natbib}

\appendix

\section{Supplementary Data Statistics}
\label{sec:data}

In Table~\ref{sumstat:personality}, we report the summary statistics of the personality estimates. We observe that all the estimates fall between 0 and 1. The standard deviations range from 0.24 to 0.28. \textit{Openness} has the highest mean value at 0.61 and \textit{conscientiousness} has the lowest mean value at 0.28.

\begin{figure*}[!htbp]
    \centering
    \includegraphics[scale = 0.21]{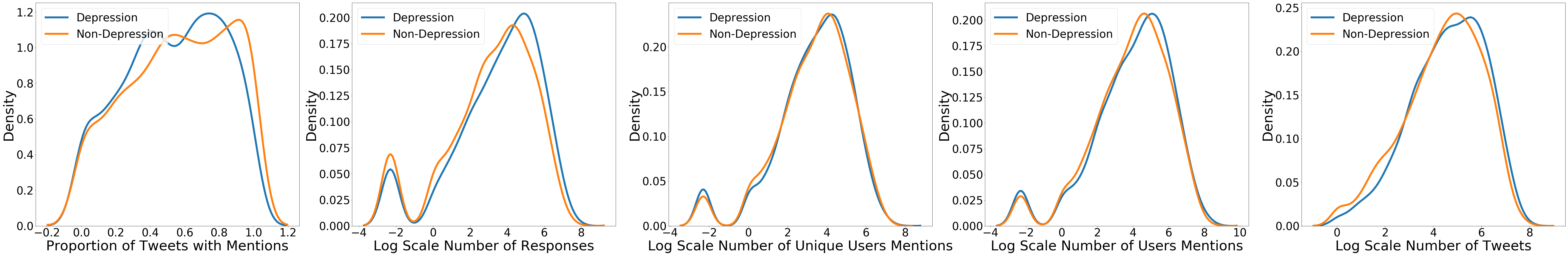}
    \caption{Density plots of proportion of tweets with mentions, log-scale number of responses, unique users mentions, users mentions, and tweets.}
    \label{fig:social_engagement}
\end{figure*}

\begin{table}[ht]
\centering
\renewcommand{\arraystretch}{1.2}
\setlength{\tabcolsep}{2pt}
\def\sym#1{\ifmmode^{#1}\else\(^{#1}\)\fi}
\begin{tabular}{llllll}
\hline\hline
            &       Count&         Min&         Max&        Mean&          SD\\
\hline
Openness    &        4697&        0.00&        1.00&        0.61&        0.28\\
Conscientiousness&        4697&        0.00&        1.00&        0.28&        0.26\\
Extraversion&        4697&        0.00&        1.00&        0.32&        0.24\\
Agreeableness&        4697&        0.00&        1.00&        0.30&        0.26\\
Neuroticism &        4697&        0.00&        1.00&        0.56&        0.28\\
\hline
\(N\)       &        4697&            &            &            &            \\
\hline\hline
\end{tabular}
\caption{Summary statistics of personality scores of users in the data set. \textit{SD} stands for standard deviation.}
\label{sumstat:personality}
\end{table}

In Table \ref{correlation}, we further report the correlation coefficients between the personality variables. We observe that \textit{extraversion} is highly correlated with \textit{conscientiousness} and \textit{agreeableness} (correlation coefficients $>$ 0.45). Meanwhile, \textit{neuroticism} is negatively correlated with \textit{openness}, \textit{conscientiousness}, and \textit{extraversion}.

\begin{table}[!htbp]
\tiny
\renewcommand{\arraystretch}{2}
\setlength{\tabcolsep}{2pt}
\def\sym#1{\ifmmode^{#1}\else\(^{#1}\)\fi}
\begin{tabular}{l*{5}{c}}
\toprule
          & Openness         &Conscientious         &Extraversion         &Agreeableness         &Neuroticism         \\
\midrule
Openness  &        1         &                  &                  &                  &                  \\
Conscientious &    0.238\sym{***}&        1         &                  &                  &                  \\
Extraversion&    0.265\sym{***}&    0.475\sym{***}&        1         &                  &                  \\
Agreeableness&  -0.0768\sym{***}&    0.441\sym{***}&    0.469\sym{***}&        1         &                  \\
Neuroticism&   -0.232\sym{***}&   -0.279\sym{***}&   -0.285\sym{***}&   0.0460\sym{**} &        1         \\
\bottomrule
\multicolumn{6}{l}{\footnotesize \sym{*} \(p<0.05\), \sym{**} \(p<0.01\), \sym{***} \(p<0.001\)}\\
\end{tabular}
\caption{Correlation between the personality variables.}
\label{correlation}
\end{table}

To better understand the difference of social media engagement between DP and ND, we add 0.1 to the number of responses, unique users mentions, users mentions, and tweets and take the logarithm. The density comparisons are shown in Figure~\ref{fig:social_engagement}. By applying the Mann-Whitney rank test, we find that except for the number of unique users mentions, other features are statistically different ($p<0.05$) between DP and ND. The users of DP post more tweets and reply more. They tend to post fewer tweets with mentions while the number the users mentioned for DP is larger, which suggests that when users of DP post tweets to interact with other users, it involves more users.

\begin{table*}[h]
    \centering
    \renewcommand{\arraystretch}{1.5}
    \begin{tabular}{|c|c|}
    \Xhline{2\arrayrulewidth}
        \multicolumn{2}{|c|}{\textbf{Model 1: Before Announcement of U.S. National Emergency}}\\
        \Xhline{2\arrayrulewidth}
        \multirow{2}{*}{School and Work} & time, day, people, anyone, thing, support, work, man, \\
        & class, today, school, week, watch, love, dog \\
        \hline
        \multirow{2}{*}{Health} & day, world, time, week, thank, today, ko,\\
        & health, hope, video, news, something, hell, job, people\\
        \hline
        \multirow{2}{*}{Trump Discussions}& people, life, time, day, work, love, trump, lot, \\
        & man, today, person, someone, thing, way, hope\\
        \hline
        \multirow{2}{*}{Cancer} & time, year, people, day, work, way, tweet, city,\\
        & cancer, head, friend, death, today, love, problem\\
        \hline
        \multirow{2}{*}{Entertainment} & love, people, time, way, song, something, day, man\\
        & nothing, everyone, tonight, year, today, game, guy\\
        \Xhline{2\arrayrulewidth}
        
        \multicolumn{2}{|c|}{\textbf{Model 2: After Announcement of U.S. National Emergency}}\\
        \Xhline{2\arrayrulewidth}
        \multirow{2}{*}{Hoarding related to COVID-19} & time, dog, people, man, everything, call, work, way, \\
        & covid, hope, news, food, night, thank, someone \\
        \hline
        \multirow{2}{*}{Quarantine and Depression} & time, work, love, day, people, something, man, thank,\\
        & thing, hope, everyone, life, quarantine, house, home\\
        \hline
        \multirow{2}{*}{Chinese News about COVID-19}& china, people, time, love, day, street, song, name,\\
        & person, trump, news, government, expert, virus, dey\\
        \hline
        \multirow{2}{*}{Depression and Anxiety during COVID-19} & people, today, time, day, covid, work, home, love,\\
        & virus, man, hope, tweet, thing, depression, everyone\\
        \hline
        \multirow{2}{*}{Work from Home} & day, people, time, love, everyone, job, quarantine,\\
        & today, year, thing, life, home, way, week, video\\
        \Xhline{2\arrayrulewidth}
        
        \multicolumn{2}{|c|}{\textbf{Model 3: State-level Topics}}\\
        \Xhline{2\arrayrulewidth}
        \multirow{2}{*}{Life Change during COVID-19} & day, today, march, morning, week, life, run, friend, \\
        & show, death, man, place, mayor, order, change \\
        \hline
        \multirow{2}{*}{Quarantine} & day, time, love, quarantine, today, ass, night, year,\\
        & thank, tomorrow, house, video, game, life, miss\\
        \hline
        \multirow{2}{*}{Hospital News}& covid, state, home, work, health, news, county, week,\\
        & hospital, testing, today, day, world, time, order\\
        \hline
        \multirow{2}{*}{Government's policy on COVID-19} & trump, people, president, time, virus, care, vote, country,\\
        & china, money, need, medium, nothing, job, everyone\\
        \hline
        \multirow{2}{*}{Family and Work} & time, people, love, man, way, job, thank,\\
        & day, work, lot, someone, something, thing, family, today\\
        \hline
    \end{tabular}
    \caption{Top 15 keywords for each topic}
    \label{tab:lda_keyword}
\end{table*}

\section{Hyper-Parameters and Other Experimental Settings}
\label{sec:experiment}

We manually select the hyper-parameters that give the best accuracy and F1 scores on the deep learning models. We use Adam optimizer with learning rate 7e-3, weight decay 1e-4 for training Attention BiLSTM. We use Adam optimizer with a learning rate of 5e-4 for training CNN. We use AdamW optimizer with a learning rate of 2e-5 for training BERT and RoBERTa, 8e-6 for training XLNet. We use the cross entropy loss for all our models during training. We use SGD optimizer with adaptive learning rate with initial learning rate as 0.1 for training SVM and logistic regression classifier. 

The experiments of the deep learning models are performed on a single NVIDIA Tesla T4 GPU. We record the models' number of parameters and their average training time per epoch on the 4,650-user training set. The BiLSTM and CNN model has 23 million and 57 million parameters, respectively (including the embedding weights). We train BiLSTM using batch size 32 and it takes 27 seconds per epoch. We train CNN using batch size 32 and it takes 39 minutes per epoch. The transformer-based models (BERT, RoBERTa, XLNet) have 110 million parameters and their speeds do not vary much. We train these models using batch size 8 and they take 1 hour 37 minutes per epoch. We train SVM, logistic regression classifier, and random forest on CPU. 

We use the \textit{ekphrasis} Python library\footnote{https://pypi.org/project/ekphrasis/} for majority of the preprocessing and use the \textit{transformers} library\footnote{https://github.com/huggingface/transformers} for building our transformer-based models.

\section{LDA Topics}
\label{sec:lda}

We use latent Drichlet allocation (LDA) provided by the \textit{gensim} library\footnote{https://pypi.org/project/gensim/} to model the topics of the tweets. To better understand the topics, we remove all the adjectives, adverbs and verbs from the text, and only keep the nouns. The number of the topics is set to be 5. In total we have trained three LDA models: one for tweets before the announcement of U.S. National Emergency, one for tweets after the announcement of U.S. National Emergency, and one for national and state level tweets between March 3rd, 2020 and May 22nd, 2020. Table~\ref{tab:lda_keyword} shows the top 15 keywords for each topic.

\end{document}